\documentclass{article}
\usepackage[top=32mm, bottom=30mm, left=25mm, right=25mm]{geometry}
\usepackage{amsmath}
\usepackage{amsthm}
\usepackage{amssymb}
\usepackage{epstopdf}
\usepackage{graphicx}
\usepackage[utf8]{inputenc}

\begin{document}
\title
{ "Killing Mie Softly": Analytic Integrals for   Resonant Scattering States }
\author{R.C. McPhedran$^1$ and Brian Stout$^2$,\\
$^1$ IPOS, School of Physics,\\
University of Sydney, 2006, Australia,\\
$^2$ Aix-Marseille Universit\'{e},  CNRS,  Ecole Centrale Marseille, Institut Fresnel,\\
 13397 Marseille, France.}
\maketitle
\begin{abstract}
We consider integrals of products of Bessel functions and of spherical Bessel functions, combined with a Gaussian factor guaranteeing convergence at infinity. Explicit representations are obtained for the integrals, building on those in the 1992 paper by McPhedran, Dawes and Scott. Attention is paid to those sums with a distributive part arising as the Gaussian tends towards a constant. Taking this limit provides explicit analytic formulae for integrals which are difficult to evaluate numerically or otherwise, and whose interpretation has been contentious both in the literature of quantum mechanics and of electromagnetic scattering.
\end{abstract}
\section{Introduction}
An important problem of long standing in quantum mechanics and nanoscale photonics concerns the behaviour of resonant modes with complex propagation constants. We will be concerned here with the fundamental problem in which those resonant modes arise in the context of scattering of electromagnetic waves by a sphere \cite{Mie1908}, but of course there has been a strong overlap between this vector problem and the scalar problem arising in quantum mechanical scattering \cite{CalderonPeierls1976,zeld61,perezeld,Calderon2010}. In both contexts, investigators have wished to employ analyses based on resonant modes, but have been worried by difficulties, real or apparent, arising from the behaviour of these modes at large distances from the scatterer, where the complex wave numbers give rise to divergent but oscillating wave functions which create problems when calculating energies or inner products. Some references to the difficulties attached to the problems in electromagnetic scattering, and to the applications which motivate their resolution, may be found in \cite{Dubovik1990}-\cite{Mulj16}.

The approach we follow here responds to a comment, or challenge, in the paper by Zel'dovich \cite{zeld61}. He notes that the absolute value of integrals associated with complex modes may be exponentially divergent at large distances, and that this cannot be remedied by exponential "kill" functions which go to zero at large distances, and which are turned off after a convergent answer is obtained.
He then suggests the use of Gaussian "kill" functions, which can be turned off gradually by letting their width increase, but which for any finite width generate a well defined answer. He then prefers to follow a different approach, based on analytic continuation and perturbation theory. We propose to follow through the Gaussian approach here in all its details for the problem of Mie scattering, a procedure for which the authors propose the shorthand "Killing Mie softly".

The approach used relies extensively on analytic results obtained by McPhedran, Dawes and Scott \cite{MDS92} (hereafter referred to as MDS), which are summarised in Section 2. Section 3 discusses the details necessary for treating the limit as the Gaussian width tends to infinity, and derives a key formula giving delta-function terms which occur in some integrals. Section 4 further extends the results of McPhedran, Dawes and Scott \cite{MDS92}, giving more general results for certain Bessel products needed in later sections. Section 5 contains a discussion of spherical Bessel function integrals needed for the treatment of complex Mie scattering modes. 
It contains a numerical illustration of the effectiveness of the Gaussian "softly killing": see Fig. \ref{figkill} and its discussion. Section 6 assembles the expressions and integrals relating to Mie theory for electric and magnetic type modes, and also establishes analytic results relevant to inner products and normalisation factors of modes.
\section{The Results of McPhedran, Dawes and Scott}
The basic integral of MDS \cite{MDS92} contains a product of Bessel functions $J$ and $Y$:
\begin{equation}
{\cal I}_{JY}(b,K,k,\eta)=\int_0^\infty x \exp(-\eta x^2) J_b(K x) Y_b(k x) d x,
\label{mds1}
\end{equation}
where $b$ is non-negative real  and $\eta$ is a  positive real. The case of integer $b$ is useful for scattering problems involving cylinders, while half-integer $b$ is useful for scattering by spheres.
This can be paired with an integral whose result is known (for example from Gradshteyn and Ryzhik \cite{GR}):
\begin{equation}
{\cal I}_{JJ}(b,K,k,\eta)=\int_0^\infty x \exp(-\eta x^2) J_b(K x) J_b(k x) d x=\frac{\exp [-(K^2+k^2)/4 \eta] I_b (K k/2 \eta)}{2\eta}.
\label{mds2}
\end{equation}

MDS derive the following reduction formula for ${\cal I}_{JY}(b,K,k,\eta)$:
\begin{eqnarray}
{\cal I}_{JY}(b,K,k,\eta&)=&\frac{-1}{2\pi \eta} \exp[-(K^2+k^2)/4 \eta]\left\{{\cal H}(b,K,k,\eta)\right.\nonumber\\
&&\left.  -2\pi\eta \exp\left[\frac{K k}{2\eta}\right]\int_0^\infty \exp(-\eta x^2) J_b(\sqrt{K k} x) Y_b(\sqrt{K k} x) d x\right\},\nonumber\\
&& 
\label{mds3}
\end{eqnarray}
or, scaling the integration variable,
\begin{eqnarray}
{\cal I}_{JY}(b,K,k,\eta&)=&\frac{-1}{2\pi \eta} \exp[-(K^2+k^2)/4 \eta]\left\{{\cal H}(b,K,k,\eta)\right.\nonumber\\
&&\left.  -\frac{2\pi\eta}{Kk} \exp\left[\frac{K k}{2\eta}\right]\int_0^\infty \exp\left(-\frac{\eta}{Kk} x^2\right) J_b( x) Y_b(x) d x\right\},\nonumber\\
&& 
\label{mds3a}
\end{eqnarray}
In (\ref{mds3})  the following finite-range integral has been introduced:
\begin{equation}
{\cal H}(b,k,K,\eta)=\int_1^{K/k} u^{(b - 1)} \exp[\frac{K k}{4\eta} (u + 1/u)] du.
\label{mds4}
\end{equation}
This integral may be expressed in terms of Scharofsky functions \cite{Peter86}, but is easily evaluated numerically. 

Making the substitution $v=1/u$ in the integral, we find the symmetry relation
\begin{equation}
{\cal H}(b,K,k,\eta)=-{\cal H}(-b,k,K,\eta).
\label{mds5}
\end{equation}
The second integral occurring in (\ref{mds3}) can be evaluated in closed form using properties of Meier $G$ functions, with
\begin{equation}
\int_0^\infty x \exp(-a x^2) J_b(x) Y_b(x) d x=\frac{1}{2\pi a} \exp[-1/(2 a)]\left[\pi \cot(\pi b) I_b\left(\frac{1}{2 a}\right)+b h_{-1,b}\left(
\frac{-1}{2 a}\right)\right].
\label{mds6}
\end{equation}
Here  $a=\eta/(K k)$, and $h_{-1,b}$ denotes an associated Bessel function\cite{Luke62}, with the expansion:
\begin{equation}
h_{-1,b}\left( \frac{-1}{2 a}\right)=-\frac{\exp[1/(2 a)]\sqrt{\pi}}{b\sin(\pi b)}\sum_{l=0}^\infty \frac{(-1/a)^l 
\Gamma (1/2+l)}{\Gamma(l-b+1)\Gamma (l+b+1)}.
\label{mds7}
\end{equation}
Note the symmetry relation
\begin{equation}
h_{-1,b}\left( a\right)=h_{-1,-b}\left(a \right).
\label{mds8}
\end{equation}

Putting these elements together, the result of MDS is
\begin{eqnarray}
{\cal I}_{JY}(b,K,k,\eta)&=&-\frac{ \exp[-(K^2+k^2)/4 \eta]}{2\pi\eta} \left\{{\cal H}(b,k,K,\eta) \right.\nonumber\\
& & \left.  -\left[\pi \cot (b \pi) I_b\left(\frac{ K k}{2\eta}\right)
+b h_{-1,b}\left( \frac{-k K}{2\eta}\right)\right]\right\}. \nonumber\\
&&
\label{mds9}
\end{eqnarray}
or
\begin{eqnarray}
{\cal I}_{JY}(b,K,k,\eta)&=&-\frac{ \exp[-(K^2+k^2)/4 \eta]}{2\pi\eta} {\cal H}(b,k,K,\eta) \nonumber\\
& &   + \frac{\exp\left[\frac{-(K^2+k^2)}{4\eta}\right]}{2\pi \eta} \left[\pi \cot (b \pi) I_b\left(\frac{ K k}{2\eta}\right)
+b h_{-1,b}\left( \frac{-k K}{2\eta}\right)\right]. \nonumber\\
&&
\label{mds10}
\end{eqnarray}
\subsection{The Case of Integer Index $b$}
It will be noted that a difficulty arises with the result (\ref{mds10}) if $b$ is an integer, since the multiplying factor in $h_{-1,b}$ involves $\sin(\pi b)$ in its denominator. This can be overcome by the use of another associated Bessel function \cite{Luke62}:
\begin{equation}
H_{\mu, \nu}(z)=h_{\mu, \nu}(z)-\frac{\sqrt{\pi}\Gamma(\mu-\nu+1)\Gamma(\mu+\nu+1)}{2^{\nu+1} \Gamma(\mu+3/2)} 
 \left[I_\nu(z)+\frac{K_\nu(z) \sin (\nu-\mu)\pi}{\pi \cos (\mu \pi)} \right].
\label{mds11}
\end{equation}
We are interested in the case $\mu=-1$, for which (\ref{mds11}) simplifies to
\begin{equation}
H_{-1, \nu}(z)=h_{-1, \nu}(z)+\frac{1}{\nu}
 \left[\frac{ \pi I_\nu(z)}{\sin (\pi \nu)}+K_\nu(z) \right].
\label{mds12}
\end{equation}
Now
\begin{eqnarray}
 & b h_{-1,b}\left(\frac{-1}{2 a}\right)+\pi \cot(\pi b)I_b\left(\frac{1}{2 a}\right)=&b H_{-1,b}\left(\frac{-1}{2 a}\right)
-\frac{\pi}{\sin (\pi b)} I_b\left(\frac{-1}{2 a}\right)     \nonumber \\
&& -K_b\left(\frac{-1}{2 a}\right) +\pi \cot (\pi b) I_b\left(\frac{1}{2 a}\right)
\label{mds131}
\end{eqnarray}
In (\ref{mds131}) we use the analytic continuation expressions \cite{NIST} (10.34):
\begin{equation}
 I_b\left(\frac{-1}{2 a}\right)= e^{\pi b i} I_b\left(\frac{1}{2 a}\right), ~~ K_b\left(\frac{-1}{2 a}\right)= e^{-\pi b i} K_b\left(\frac{1}{2 a}\right)
 -\pi i  I_b\left(\frac{1}{2 a}\right).
\label{mds132}
\end{equation}
The latter expression in (\ref{mds132}) makes evident the branch-cut behaviour of the Macdonald function near the negative real axis.
We collect all the terms in $I_b(1/2 a)$ on the right-hand side of (\ref{mds131}), which cancel, leaving
\begin{equation}
b h_{-1,b}\left(\frac{-1}{2 a}\right)+\pi \cot(\pi b)I_b\left(\frac{1}{2 a}\right)=b H_{-1,b}\left(\frac{-1}{2 a}\right) -e^{-i\pi b} K_b\left(\frac{1}{2 a}.\right) 
\label{mds13}
\end{equation}
Hence, (\ref{mds6}) becomes for $n$ an integer:
\begin{equation}
\int_0^\infty x \exp(-a x^2) J_n(x) Y_n(x) d x=\frac{1}{2\pi a} \exp[-1/(2 a)]\left[  n H_{-1,n}\left(\frac{-1}{2 a}\right)+(-1)^{n+1} K_n\left(\frac{1}{2a}\right)\right].
\label{mds14}
\end{equation}
The following terminating series can be used for $H_{-1,n}$:
\begin{align}
& H_{-1,n}\left(\frac{-1}{2 a}\right)=-2 a \exp \left(\frac{1}{2 a}\right) \left[ 1+\frac{2 a}{3} (1-n^2)+\frac{4 a^2}{15} (1-n^2)(4-n^2)\right. \label{mds15} \\
&\left. +\frac{8 a^3}{105} (1-n^2)(4-n^2)(9-n^2)+\frac{16 a^4}{945} (1-n^2)(4-n^2)(9-n^2)(16-n^2)+\ldots \right].\nonumber
\end{align}

Particular cases of equation (\ref{mds14}) are:
\begin{equation}
\int_0^\infty x \exp(-a x^2) J_0(x) Y_0(x) d x=\frac{1}{2\pi a} \exp[-1/(2 a)]\left[-K_0\left( \frac{1}{2 a}\right)\right],
\label{mds16}
\end{equation}
\begin{equation}
\int_0^\infty x \exp(-a x^2) J_1(x) Y_1(x) d x=\frac{-1}{\pi}+\frac{1}{2\pi a} \exp[-1/(2 a)]\left[K_1\left( \frac{1}{2 a}\right)\right],
\label{mds17}
\end{equation}
\begin{equation}
\int_0^\infty x \exp(-a x^2) J_2(x) Y_2(x) d x=\frac{-2}{\pi}(1-2 a)+\frac{1}{2\pi a} \exp[-1/(2 a)]\left[-K_2\left( \frac{1}{2 a}\right)\right]
\label{mds18}
\end{equation}
and 
\begin{equation}
\int_0^\infty x \exp(-a x^2) J_3(x) Y_3(x) d x=\frac{-3}{\pi}(1-\frac{16}{3} a+\frac{32}{3} a^2)+\frac{1}{2\pi a} \exp[-1/(2 a)]\left[ K_3\left( \frac{1}{2 a}\right)\right].
\label{mds18bis}
\end{equation}
As $a\downarrow 0$, we have the following limiting behaviour of these:
\begin{equation}
\int_0^\infty x \exp(-a x^2) J_0(x) Y_0(x) d x\rightarrow \frac{-1}{2\sqrt{\pi a}} \exp \left( \frac{-1}{a}\right),
\label{mds16a}
\end{equation}
\begin{equation}
\int_0^\infty x \exp(-a x^2) J_1(x) Y_1(x) d x\rightarrow \frac{-1}{\pi}+ \frac{1}{2\sqrt{\pi a}} \exp \left( \frac{-1}{a}\right),
\label{mds17a}
\end{equation}
\begin{equation}
\int_0^\infty x \exp(-a x^2) J_2(x) Y_2(x) d x\rightarrow \frac{-2}{\pi}(1-2 a)- \frac{1}{2\sqrt{\pi a}} \exp \left( \frac{-1}{a}\right),
\label{mds18a}
\end{equation}
and
\begin{equation}
\int_0^\infty x \exp(-a x^2) J_3(x) Y_3(x) d x\rightarrow \frac{-3}{\pi}(1-\frac{16}{3} a+\frac{32}{3} a^2)+\frac{1}{2\sqrt{\pi a}} \exp \left( \frac{-1}{a}\right),
\label{mds18bisa}
\end{equation}

The first of these tends to zero  as  the exponential form $\exp(-1/a)$, while succeeding expressions  tend to  $-n/\pi$.

\section{Some  Limits of Integrals}
Let us first consider the limit as $\eta\rightarrow 0$ of ${\cal I}_{JJ}(b,K,k,\eta)$, in the sense of generalised functions. This evaluation is done in a way similar to that followed by Lekner, Appendix B, Chapter 4 \cite{Lekner18}. We use the large argument expansion of $I_{b}$:
\begin{equation}
I_{b}(x)\sim \frac{\exp (x)}{\sqrt{2 \pi x}} \left(1-\frac{(b-1/2) (b+1/2)}{2 x}\right)+\exp(i(b+1/2)\pi)\frac{\exp (-x)}{\sqrt{2 \pi x}} \left(1+\frac{(b-1/2) (b+1/2)}{2 x}\right),
\label{nr16}
\end{equation}
where the first omitted term is of order $1/x^{5/2}$.
With $a=1/\sqrt{\eta}$ tending to infinity, we have in (\ref{mds2}) the term
\begin{eqnarray}
&&\frac{1}{2} a^2 \exp[-(K^2+k^2) a^2/4]  I_{b}\left(\frac{K k a^2}{2 }\right) \sim  \left\{\frac{1}{2 } 
a^2 \exp[-(K^2+k^2) a^2/4] \right\} \nonumber \\
&& \left[ \frac{\exp (K k a^2/2)}{\sqrt{ \pi K k a^2}}   \left(1-\frac{(b-1/2) (b+1/2)}{K k a^2}\right)+\exp(i(b+1/2)\pi) \frac{\exp (-K k a^2/2)}{\sqrt{ \pi K k a^2}}   \left(1+\frac{(b-1/2) (b+1/2)}{K k a^2}\right) \right] . \nonumber \\
& &  
\label{nr17}
\end{eqnarray}
We complete the squares of the exponential arguments in (\ref{nr17}) to obtain:
\begin{eqnarray}
&&\frac{1}{2} a^2 \exp[-(K^2+k^2) a^2/4]  I_{b}\left(\frac{K k a^2}{2 }\right) \sim  \left\{\frac{1}{2 } 
a^2 \right\} \nonumber \\
&& \left[ \frac{\exp (-(K-k)^2 a^2/4)}{\sqrt{ \pi K k a^2}}   \left(1-\frac{(b-1/2) (b+1/2)}{K k a^2}\right)+\right.\nonumber \\
& & \left.\exp(i(b+1/2)\pi) \frac{\exp (-(K+ k)^2 a^2/4)}{\sqrt{ \pi K k a^2}}   \left(1+\frac{(b-1/2) (b+1/2)}{K k a^2}\right) \right] . \nonumber \\
& &  
\label{nr18}
\end{eqnarray}
We recognise the terms in (\ref{nr18}) corresponding to $\delta$ functions:
\begin{equation}
\frac{a}{2 \sqrt{\pi}}\exp (-a^2 x^2/4)\rightarrow \delta (x), ~{\rm for}~a\rightarrow\infty.
\label{nr19}
\end{equation}
Hence
\begin{equation}
\lim_{\eta\rightarrow 0} {\cal I}_{JJ}(b,K,k,\eta)=\frac{1}{ \sqrt{K k}} \delta (K-k).
\label{nr20}
\end{equation}
A term $\exp(i(b+1/2)\pi) \delta (k+K)$ in (\ref{nr20}) has been  be omitted under the assumption that  $K+k=0$ can be excluded.

Note that the asymptotic expansion of the integrand in (\ref{mds2}) has, with $\eta=0$, the leading term
\begin{equation}
\frac{2}{\pi  \sqrt{K k}} (\cos [(K-k) x] \pm \cos [(K+k) x]).
\label{mds2a}
\end{equation}
This then makes the origin of the full expression (\ref{nr20}) (with both $\delta$ functions included) evident: it is associated with the behaviour of the integrand near the  upper limit of integration. We thus expect similar delta function terms to appear whenever the asymptotic development of the integral generates either a $\cos(K x) \cos(k x)$ or a $\sin(K x) \sin(k x)$ variation. Both these products oscillate ever more rapidly about mean zero as $x\rightarrow \infty$ unless $k=\pm K$ (in which cases they have mean 1/2 or -1/2).

We now turn our attention to the behaviour as $\eta\rightarrow 0$ of 
\begin{equation}
{\hat {\cal H}}(b,k,K,\eta)=-\frac{\exp[-(K^2+k^2)/4 \eta]}{2\pi \eta} \int_1^{K/k} u^{(b - 1)} \exp[\frac{K k}{4\eta} (u + 1/u)] du.
\label{nr21}
\end{equation}
The integrand in (\ref{nr21}) increases monotonically as $u$ increases towards the upper limit. We analyse the asymptotic behaviour of the function ${\hat {\cal H}}(b,k,K,\eta)$ by defining $v=K/k-u$, and bringing all terms on the right-hand side of (\ref{nr21}) into a single exponential term. We then expand the argument of the exponential in powers of $v$. Retaining the zeroth, first and second powers of $v$  we find the  approximation:
\begin{eqnarray}
{\hat {\cal H}}(b,k,K,\eta)&\approx &-\frac{1}{2\pi \eta}\int_0^{K/k-1} dv  \exp\left\{(b-1) \log\left[\frac{K}{k}\right]] +\right.  \nonumber\\
 && \left. v\left[\frac{ k^3 - k K^2 + 4 k \eta - 4 b k \eta}{4 K \eta} \right]+ 
 v^2 \left[\frac{k^4 + 2 k^2 \eta - 2 b k^2 \eta}{4 K^2 \eta} \right]\right\} .\nonumber \\
&& 
\label{nr22}
\end{eqnarray}
Mathematica gives an expression for the integral in (\ref{nr22}) involving a combination of two imaginary error functions:
\begin{eqnarray}
&&{\hat {\cal H}}(b,k,K,\eta)\approx -\frac{1}{2\pi \eta} \times \nonumber \\
&&  \hspace{-1cm} \frac{\sqrt{\pi } \sqrt{\eta } \left(\frac{K}{k}\right)^b \left(\text{erfi}\left(\frac{4
   (b-1) \eta  k-K \left(8 (b-1) \eta +K^2\right)-2 k^3+3 k^2 K}{4 \sqrt{\eta } K
   \sqrt{k^2-2 (b-1) \eta }}\right)-\text{erfi}\left(\frac{-4 b \eta +4 \eta +k^2-K^2}{4
   \sqrt{\eta } \sqrt{k^2-2 (b-1) \eta }}\right)\right) \exp \left(\frac{\left(4 (b-1)
   \eta -k^2+K^2\right)^2}{16 \eta  \left(2 (b-1) \eta -k^2\right)}\right)}{\sqrt{k^2-2
   (b-1) \eta }} \nonumber \\
   &&
\label{nr23}
\end{eqnarray}
Of these two terms, the simpler expression is exponentially larger than the more complicated term. Neglecting the latter, and expanding the result as a power series  in $\eta$, the first two terms give
\begin{equation}
{\hat {\cal H}}(b,k,K,\eta)\approx  \frac{2 \left(\frac{K}{k}\right)^b}{\pi 
   \left(k^2-K^2\right)}+\frac{8 \eta  \left(b k^2-b K^2+k^2+K^2\right) \left(\frac{K}{k}\right)^b}{\pi 
   \left(k^2-K^2\right)^3} +O(\eta^2).
\label{nr24}
\end{equation}
This result, for $\eta \rightarrow 0$, is consistent with the result which comes from applying Watson's expression from p.134 of "A Treatise on the Theory of Bessel Functions" \cite{Watson80}:
\begin{equation}
\int_0^\infty   x   J_b(K x) Y_b(k x) d x=\frac{2 \left(\frac{K}{k}\right)^b}{\pi \left(k^2-K^2\right)},
 \label{nr25}
 \end{equation}
where the contribution is solely from the lower limit of the integral, with the oscillating contribution from the upper limit being set (or averaged) to zero. This last comment is made in the context of  the theory of distributions, not well established at the time Watson wrote his treatise. 
\subsection{The Series $h_{-1,b}$}
We begin with the expansion (\ref{mds7}):
\begin{equation}
h_{-1,b}\left( \frac{-1}{2 a}\right)=-\frac{\exp[1/(2 a)]\sqrt{\pi}}{b\sin(\pi b)}\sum_{l=0}^\infty \frac{(-1/a)^l \Gamma (1/2+l)}{\Gamma(l-b+1)
\Gamma (l+b+1)}.
\label{mds7bis}
\end{equation}
We expand for large $l$ the expression:
\begin{eqnarray}
\label{hmexp1}
&&-\frac{\sqrt{\pi}}{b\sin(\pi b)}\frac{(-1/a)^l \Gamma (1/2+l) l!}{\Gamma(l-b+1)
\Gamma (l+b+1)} \sqrt{l+1}=  -\frac{\sqrt{\pi}}{b\sin(\pi b)} \nonumber\\
&& \left(\frac{-1}{a}\right)^l  \frac{1}{l! \sqrt{l+1}} \left(1+\frac{(3-8 b^2)}{8 l}+\frac{(64 b^4+16 b^2-23)}{128 l^2}+O\left(\frac{1}{l^3}\right)\right).
\end{eqnarray}
In the region of slow convergence of the sum (\ref{mds7bis}), i.e when $a<<1$, it will then be dominated by the following approximation,
for which an exact integral form is available [Prudnikov, Vol.1, 5.2.8.10]:
\begin{equation}
\sum_{k=0}^\infty \frac{x^k}{k!\sqrt{k+1}}=\frac{2}{\sqrt{\pi}} \int_0^\infty \exp (x e^{-t^2}-t^2) dt.
\label{hmexp2}
\end{equation}
Note that here we have corrected a typographical error in the upper limit of the integral in (\ref{hmexp2}).

It is possible to vary the expansion (\ref{hmexp1}), for example by replacing the term $\sqrt{l+1}$ by $\sqrt{l+\alpha}$, where $\alpha$ is a free parameter. One way of choosing $\alpha$ is to make the coefficient of $1/k$ in (\ref{hmexp1}) go to zero. To achieve this,
the choice of $\alpha$ required is
\begin{equation}
\alpha=2 b^2 + 1/4 .
\label{alphachoice}
\end{equation}

We can generalise the sum and integral in (\ref{hmexp2}) as follows:
\begin{equation}
\sum_{k=0}^\infty \frac{x^k}{k!\sqrt{k+\alpha}}=\frac{2}{\sqrt{\pi}} \int_0^\infty \exp (x e^{-t^2}-\alpha t^2) dt.
\label{hmexp3}
\end{equation}
The derivation of (\ref{hmexp3}) is simple:
\begin{eqnarray}
\frac{2}{\sqrt{\pi}} \int_0^\infty \exp (x e^{-t^2}-\alpha t^2) dt &=& \frac{2}{\sqrt{\pi}} \int_0^\infty
\sum_{k=0}^\infty \frac{x^k  \exp (-k t^2)}{k!}  \exp (-\alpha t^2) dt  \nonumber \\
& &=  \sum_{k=0}^\infty \frac{x^k}{k!\sqrt{k+\alpha}},
\label{hmexp4}
\end{eqnarray}
where to get the final result the two Gaussian terms have been combined and integrated over. Note that this equality holds for all $x>0$,
but in the case of interest to us ($x<0$) a problem can arise. To see where the problem comes from, we look for the vanishing of the
derivative of the integrand with respect to $t$ in (\ref{hmexp4}):
\begin{equation}
2 t (x e^{-t^2}+\alpha )=0 ~{\rm or} ~t=0,~ x e^{-t^2}=-\alpha.
\label{hmexp5}
\end{equation}
The second possibility in (\ref{hmexp5}) has solutions on the real axis when $x<0$ and $t^2=-\log (\alpha/|x|)$. Thus, if the positive quantity $\alpha <-x$, there will be two maxima, with a minimum at $t=0$. In that case, the two peaks correspond to truncated Gaussians, and the equality (\ref{hmexp4}) breaks down.

Somewhat counterintuitively, there is a simple way of overcoming this difficulty.  We simply split the integrand into two parts, one even and the other odd in the variable $x$. It is easily shown that the integrand in each part has a single derivative zero at $t=0$, with the following results then holding for all $x$, positive or negative:
\begin{equation}
\sum_{k=0}^\infty \frac{x^{2 k}}{(2 k)!\sqrt{2 k+\alpha}}=\frac{2}{\sqrt{\pi}} \int_0^\infty \cosh (x e^{-t^2})\exp (-\alpha t^2) dt,
\label{hmexp6}
\end{equation}
and
\begin{equation}
\sum_{k=0}^\infty \frac{x^{2 k+1}}{(2 k+1)!\sqrt{2 k+1+\alpha}}=\frac{2}{\sqrt{\pi}} \int_0^\infty \sinh (x e^{-t^2}) \exp (-\alpha t^2) dt.
\label{hmexp7}
\end{equation}
These results have been confirmed numerically, a task  easily carried out for $\alpha$ not small.

\begin{figure}[tbh]
\includegraphics[width=6cm]{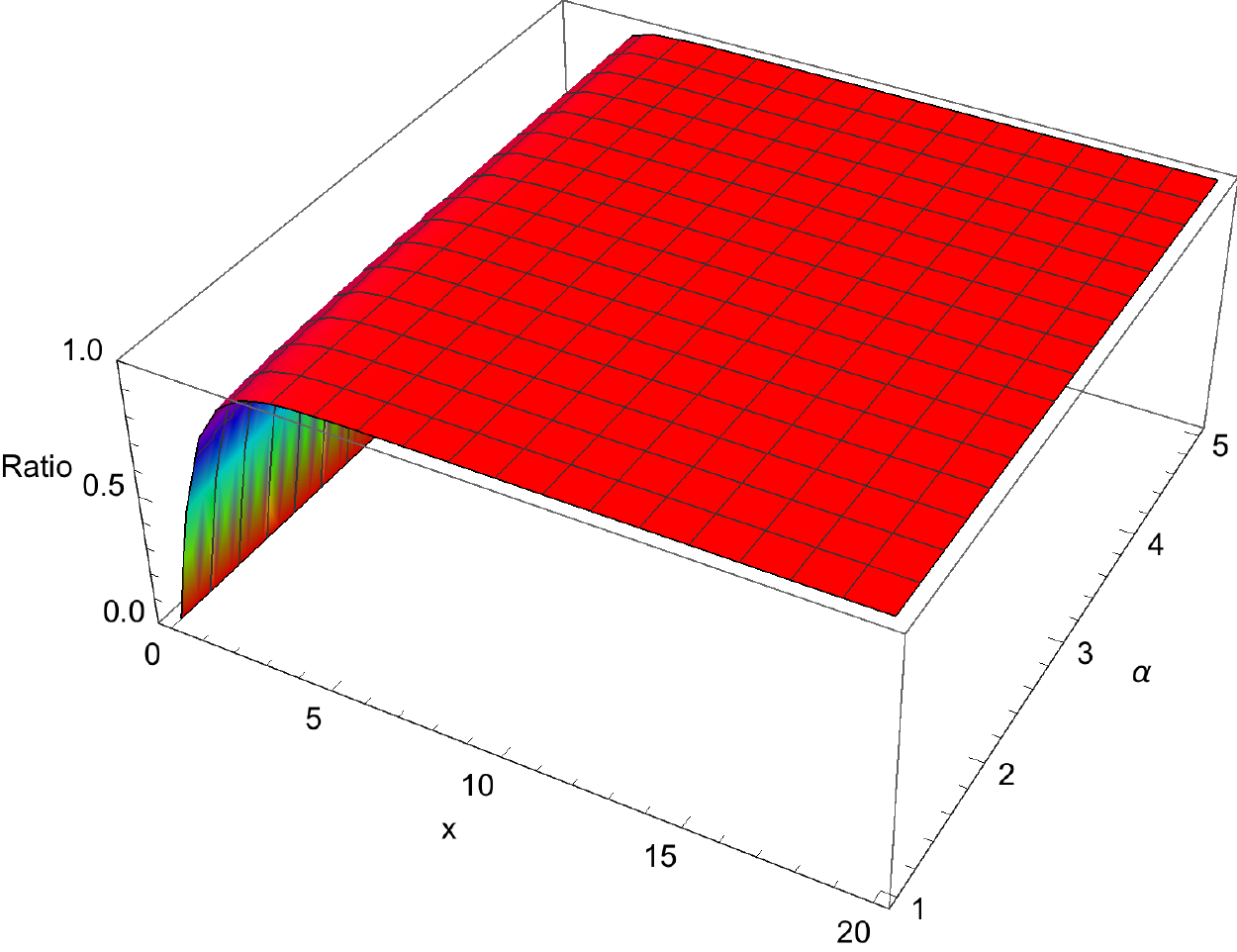}~~\includegraphics[width=6cm]{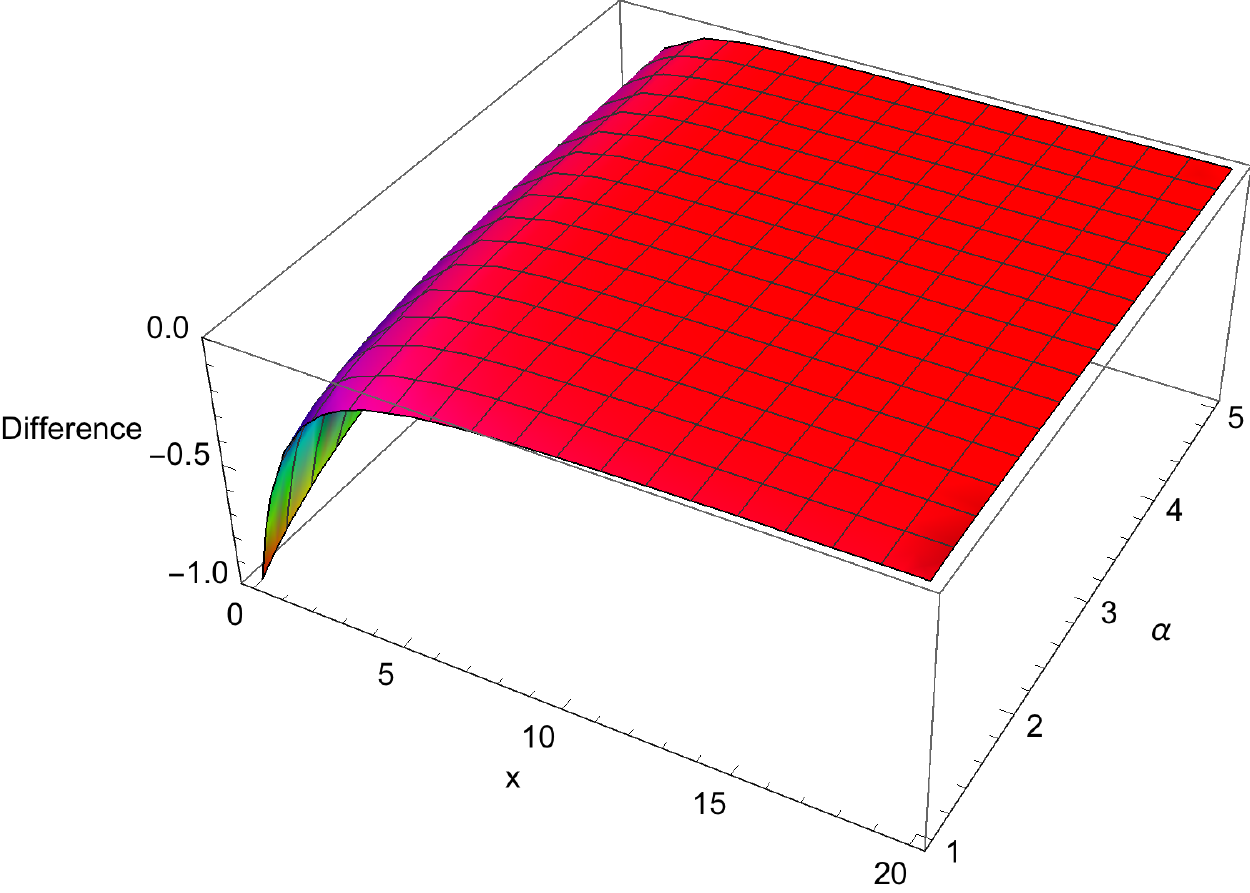}
\caption{The ratio (left) and the difference (right) of the functions (\ref{hmexp6}) and (\ref{hmexp7}) as a function of $x$ and $\alpha$.}
\label{fig-limits1}
\end{figure}

The numerical results in Fig. \ref{fig-limits1} show that the ratio of the sums or integrals in (\ref{hmexp6}) and (\ref{hmexp7}) tends towards unity as $x$ increases, irrespective of the value of $\alpha$. This may be readily understood from the ratio of the integrands,
$\tanh (x e^{-t^2})$, which increases towards unity as $x$ increases, for fixed $t$. The difference of the sums or integrals tends towards zero in its magnitude as $x$ increases.

\begin{figure}[tbh]
\includegraphics[width=9cm]{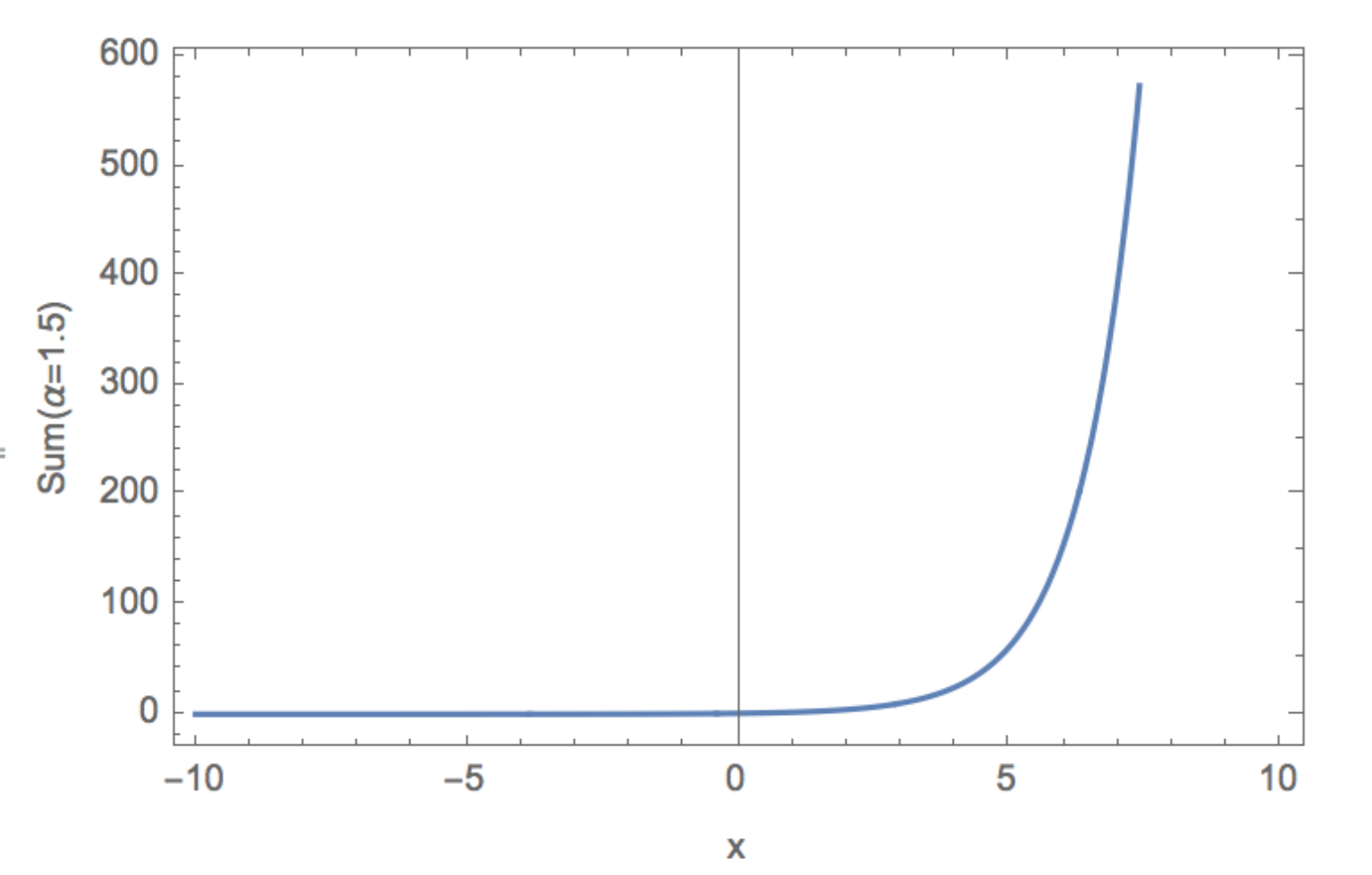}
\caption{The sum of the functions (\ref{hmexp6}) and (\ref{hmexp7}) as a function of $x$ for $\alpha=1.5$.}
\label{fig-limits2}
\end{figure}

Fig. \ref{fig-limits2} shows the sum of the functions (\ref{hmexp6}) and (\ref{hmexp7}) as a function of $x$ for $\alpha=1.5$, for both positive and negative $x$. While the sum increases exponentially for $x$ increasing and positive, it remains small for negative $x$, and decreases towards zero as $x$ grows more negative.

\begin{figure}[tbh]
\includegraphics[width=5cm]{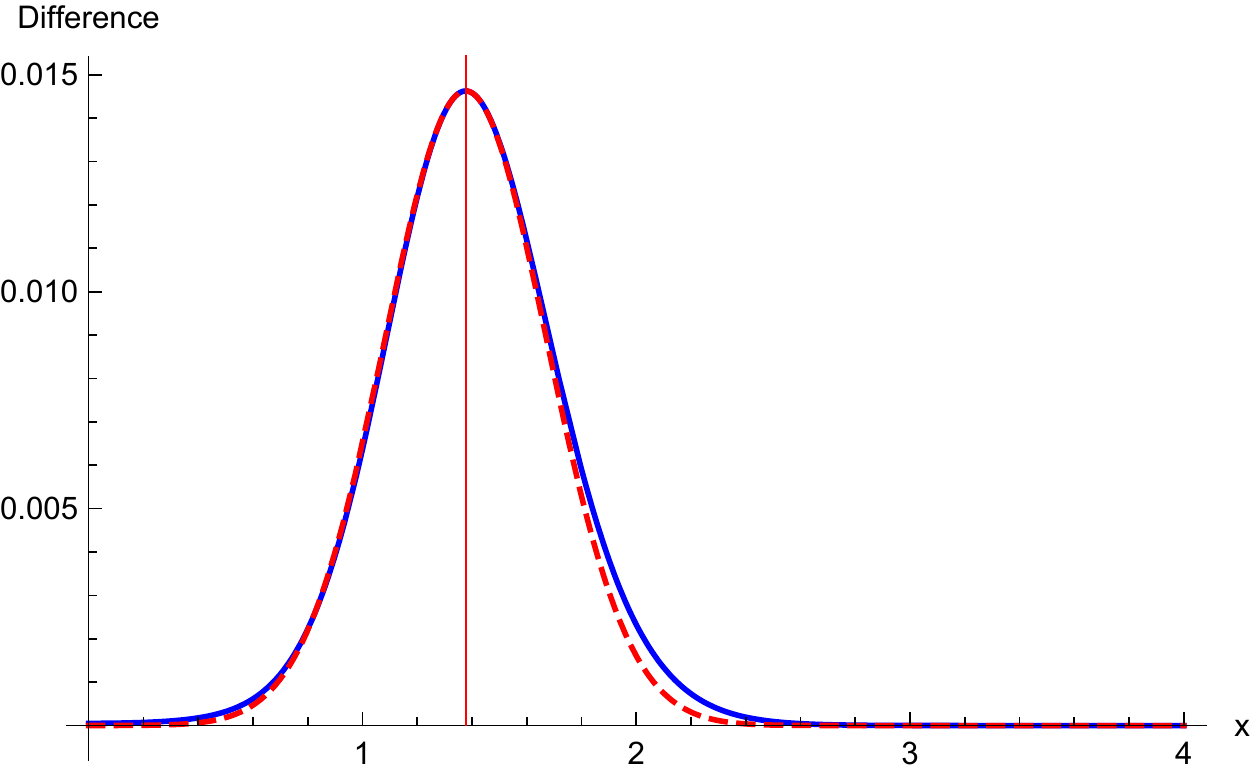}~~\includegraphics[width=5cm]{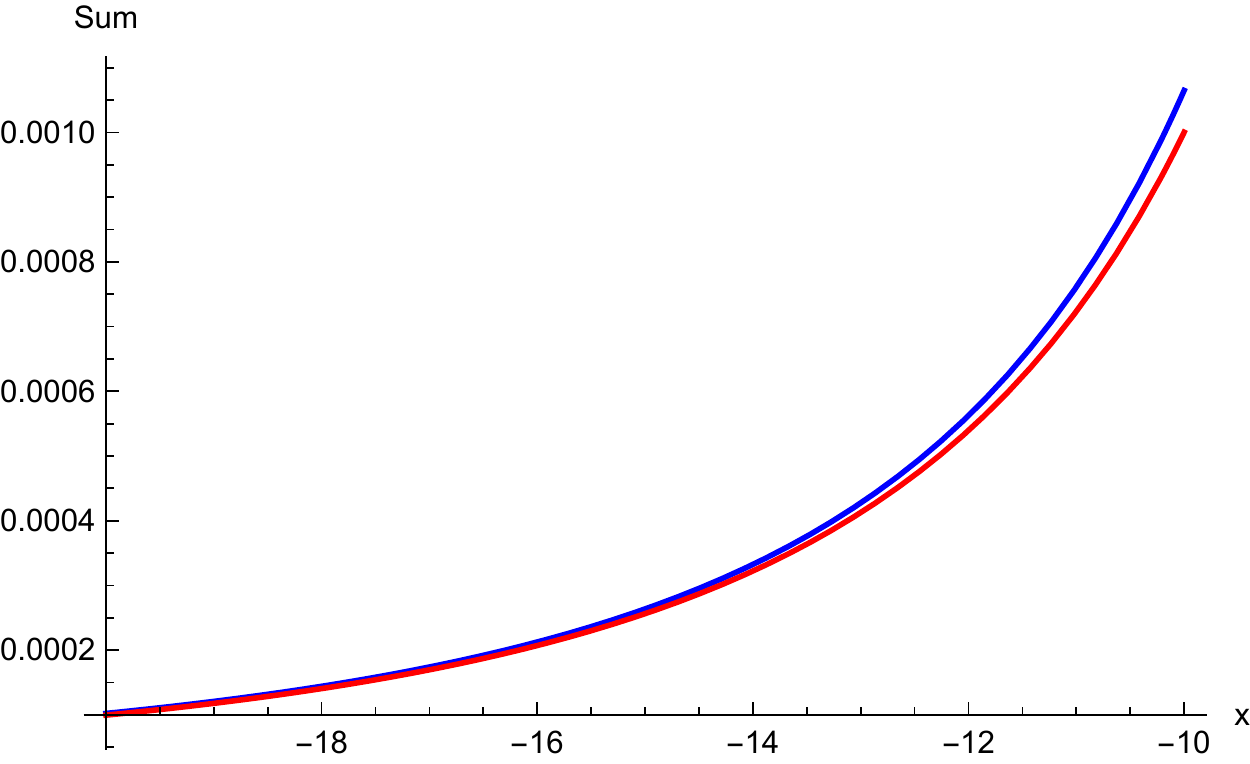}
\caption{(Left) The difference of the functions (\ref{hmexp6}) and (\ref{hmexp7}) in $x>0$ as a function of $x$ for $\alpha=1.5$ (blue line) and its Gaussian approximation (\ref{hmexp10}) (red dashed line). (Right) The sum of the functions (\ref{hmexp6}) and (\ref{hmexp7}) in $x<0$ as a function of $x$ for $\alpha=1.5$ (blue line) and the asymptotic estimate (\ref{hmexp11}) (red line). }
\label{fig-limits3}
\end{figure}

To obtain an asymptotic form in the region $x<<0$, we consider the difference function between (\ref{hmexp6}) and (\ref{hmexp7}) 
in $x>0$. This function is
\begin{equation}
\frac{2}{\sqrt{\pi}} \int_0^\infty \exp (x e^{-t^2}-\alpha t^2) dt ,
\label{hmexp8}
\end{equation}
and the integrand has its maximum with respect to variation of $t$ when
\begin{equation}
t=t_m=\sqrt{\log\left( \frac{x}{\alpha}\right)}.
\label{hmexp9}
\end{equation} 
Expanding about $t=t_m$ and retaining the Gaussian components, we obtain the approximation
\begin{equation}
\frac{2}{\sqrt{\pi}} e^{-\alpha [1+\log(x/\alpha)]} e^{-2\alpha \log(x/\alpha)(t-t_m)^2}.
\label{hmexp10}
\end{equation}
Integrating over the Gaussian approximation we obtain the following estimate for the difference of  (\ref{hmexp6}) and (\ref{hmexp7})
in $x>0$:
\begin{equation}
\frac{2 \exp -[\alpha (1+\log (x/\alpha))]}{\sqrt{2\alpha \log(x/\alpha)}}.
\label{hmexp11}
\end{equation} 
Fig. \ref{fig-limits3} illustrates the accuracy of the Gaussian approximation (\ref{hmexp10}) and the resulting estimate  (\ref{hmexp11}).
From the estimate (\ref{hmexp11}) we see that the limit as $x\rightarrow -\infty$ of the sum of the functions (\ref{hmexp6}) and (\ref{hmexp7}) is zero, for any $\alpha>0$. This is also evident from the plot on the right in Fig. \ref{fig-limits1}.

\section{Extending the Results of McPhedran, Dawes and Scott}

The extended set of results comes from use of the interrelations between Bessel functions:
\begin{equation}
Y_b(x)=\frac{J_b(x) \cos (b \pi)-J_{-b}(x)}{\sin (b \pi)},~~J_b(x)=\frac{Y_{-b}(x) \cos (b \pi)-Y_{b}(x)}{\sin (b \pi)},
\label{snr1}
\end{equation}
together with the symmetry relationships (\ref{mds5}) and (\ref{mds8}).

The first of these is for:
\begin{equation}
{\cal I}_{JJm}(b,K,k,\eta)=\int_0^\infty x \exp(-\eta x^2) J_b(K x) J_{-b}(k x) d x,
\label{snr2}
\end{equation}
for which
\begin{equation}
{\cal I}_{JJm}(b,K,k,\eta)=\frac{ \sin (b \pi)  \exp[-(K^2+k^2)/4 \eta]}{2\pi \eta} \left[{\cal H}(b,k,K,\eta)-b  h_{-1,b}\left( \frac{-k K}{2\eta}\right)\right].
\label{snr3}
\end{equation}
The second comes from replacing $b$ by $-b$ in (\ref{snr2}) and (\ref{snr3}):
\begin{equation}
{\cal I}_{JmJ}(b,K,k,\eta)=\int_0^\infty x \exp(-\eta x^2) J_{-b}(K x) J_{b}(k x) d x,
\label{snr4}
\end{equation}
and
\begin{equation}
{\cal I}_{JmJ}(b,K,k,\eta)=\frac{ \sin (b \pi)  \exp[-(K^2+k^2)/4 \eta]}{2\pi \eta} \left[{\cal H}(b,K,k,\eta)-b  h_{-1,b}\left( \frac{-k K}{2\eta}\right)\right].
\label{snr5}
\end{equation}
Hence,
\begin{eqnarray}
{\cal I}_{JJm}(b,K,k,\eta)+{\cal I}_{JmJ}(b,K,k,\eta)&=&\frac{ \sin (b \pi)  \exp[-(K^2+k^2)/4 \eta]}{2\pi \eta} \left[{\cal H}(b,K,k,\eta)
\right.\nonumber \\
 & &  \left.
+{\cal H}(b,k,K,\eta)-2 b  h_{-1,b}\left( \frac{-k K}{2\eta}\right)\right],
\label{snr6}
\end{eqnarray}
and
\begin{equation}
{\cal I}_{JJm}(b,K,k,\eta)-{\cal I}_{JmJ}(b,K,k,\eta)=\frac{ \sin (b \pi)  \exp[-(K^2+k^2)/4 \eta]}{2\pi \eta} \left[{\cal H}(b,K,k,\eta)-{\cal H}(b,K,k,\eta)\right].
\label{nr7}
\end{equation}

The third evaluation concerns:
\begin{equation}
{\cal I}_{YY}(b,K,k,\eta)=\int_0^\infty x \exp(-\eta x^2) Y_b(K x) Y_{b}(k x) d x.
\label{snr8}
\end{equation}
The right-hand side is expanded using (\ref{snr1})  twice, giving the expressions
\begin{eqnarray}
{\cal I}_{YY}(b,K,k,\eta)&=&\frac{1}{\sin^2(b \pi)}\left[\cos^2(b \pi) {\cal I}_{JJ}(b,k,K,\eta)+{\cal I}_{JJ}(-b,k,K,\eta)\right. \nonumber \\
& &\left. -\cos (b \pi) ({\cal I}_{JJm}(b,K,k,\eta)+{\cal I}_{JmJ}(b,K,k,\eta)\right] ,
\label{snr9}
\end{eqnarray}
and so
\begin{eqnarray}
{\cal I}_{YY}(b,K,k,\eta)&=&\frac{ \exp[-(K^2+k^2)/4 \eta]}{2 \eta} \left[ \cot^2 (b \pi) I_b\left(\frac{K k}{2 \eta}\right) +I_{-b}\left(\frac{K k}{2 \eta}\right)/\sin^2 (b\pi)
\right. \nonumber\\
& & \left.  -\frac{\cot (b \pi)}{\pi} \left( {\cal H}(b,K,k,\eta)+{\cal H}(b,k,K,\eta)-2 b  h_{-1,b}\left( \frac{-k K}{2\eta}\right) \right)\right].
\label{snr10}
\end{eqnarray}

We can also expand the right-hand side of (\ref{snr8}) using (\ref{snr1}) only once. Solving, we find
\begin{eqnarray}
{\cal I}_{JmY}(b,K,k,\eta) &=& \int_0^\infty x \exp(-\eta x^2) J_{-b}(K x) Y_{b}(k x) d x \nonumber \\
 &=&  \cos( b\pi) {\cal I}_{JY}(b,K,k,\eta)-\sin (b \pi) {\cal I}_{YY}(b,K,k,\eta) . 
\label{snr11}
\end{eqnarray}
As well,
\begin{eqnarray}
{\cal I}_{JYm}(b,K,k,\eta) &=& \int_0^\infty x \exp(-\eta x^2) J_{b}(K x) Y_{-b}(k x) d x \nonumber \\
 &=&  \cos( b\pi) {\cal I}_{JY}(-b,K,k,\eta)+\sin (b \pi) {\cal I}_{YY}(-b,K,k,\eta). 
\label{snr12}
\end{eqnarray}
\section{Integrals over Spherical Bessel Functions}
The results in the three preceding sections can be extended to spherical Bessel functions, using the substitutions:
\begin{equation}
j_n(z)=\sqrt{\frac{\pi}{2 z}} J_{n+1/2}(z)=(-1)^n \sqrt{\frac{\pi}{2 z}} Y_{-n-1/2}(z),
\label{snr13}
\end{equation}
and 
\begin{equation}
y_n(z)=\sqrt{\frac{\pi}{2 z}} Y_{n+1/2}(z)=(-1)^{n+1} \sqrt{\frac{\pi}{2 z}} J_{-n-1/2}(z).
\label{snr14}
\end{equation}
 
 The first integral we consider is 
 \begin{eqnarray} 
 {\cal I}_{j j}(n,K,k,\eta) &=& \int_0^\infty x^2 \exp(-\eta x^2) j_{n}(K x) j_{n}(k x) d x \nonumber \\
 &=& \frac{\pi}{2 \sqrt{K k} }\frac{ \exp[-(K^2+k^2)/4 \eta]}{2 \eta} I_{n+1/2}\left(\frac{K k}{2 \eta}\right).
\label{snr15}
\end{eqnarray}
Using the argument from Section 2 expressed in equations  (\ref{nr17}-\ref{nr19}), we have
\begin{equation}
\lim_{\eta\rightarrow 0} \left\{ \frac{1}{2 \eta}\exp\left[\frac{-(K^2+k^2)}{4\eta}\right] I_b\left( \frac{K k}{2\eta}\right) \right\}
=\frac{1}{\sqrt{K k}} \delta (K-k).
\label{snr16}
\end{equation}
Hence,
\begin{equation}
\lim_{\eta\rightarrow 0}  {\cal I}_{j j}(n,K,k,\eta)=\frac{\pi}{2 K k} \delta (K-k).
\label{snr17}
\end{equation}

The second integral to be considered is (\ref{mds10}), which, for $b=n+1/2$ gives
 \begin{eqnarray} 
 {\cal I}_{j y}(n,K,k,\eta) &=& \int_0^\infty x^2 \exp(-\eta x^2) j_{n}(K x) y_{n}(k x) d x \nonumber \\
  &=& \frac{\pi}{2}\left\{ \frac{\exp [-(K^2+k^2)/(4 \eta)]}{2\pi \eta} \left[ -{\cal H}(n+1/2,k,K,\eta) +(n+1/2) h_{-1,n+1/2}\left(\frac{-K k}{2\eta}\right) \right]\right\}. \nonumber\\
  &&
\label{snr18}
\end{eqnarray}
Now, from (\ref{nr24}),
\begin{equation}
\lim_{\eta\rightarrow 0}  \frac{\pi}{2}\left\{ \frac{\exp [-(K^2+k^2)/(4 \eta)]}{2\pi \eta} \left[ -{\cal H}(n+1/2,k,K,\eta)\right]\right\}  =\frac{ \left(\frac{K}{k}\right)^b}{ \left(k^2-K^2\right)}.
\label{snr18a}
\end{equation}
Also, from (\ref{mds7bis}),
\begin{equation}
h_{-1,n+1/2}\left(\frac{-K k}{2\eta}\right)=\frac{-\exp\left(\frac{K k}{2\eta}\right)\sqrt{\pi}}{(n+1/2) \sin(\pi(n+1/2))}
\sum_{l=0}^\infty\frac{\Gamma(1/2+l)}{\Gamma(l-n+1/2)\Gamma(l+n+3/2)} \left(\frac{-K k}{\eta}\right)^l.
\label{snr19}
\end{equation}
Hence,using the asymptotic estimate (\ref{hmexp11}) with $\alpha=2 n^2+2 n+1/2$, the contribution of this term in (\ref{snr18}) goes as
\begin{eqnarray}
 &&\frac{\pi}{2}\left\{ \frac{\exp [-(K^2+k^2)/(4 \eta)]}{2\pi \eta} (n+1/2) h_{-1,n+1/2}\left(\frac{-K k}{2\eta}\right) \right\}\approx \nonumber \\
 && (-1)^{n+1} \pi^{1/2} \frac{\exp [-(K-k)^2/(4 \eta)]}{2 \eta} \frac{ \exp -[\alpha (1+\log (K k/(\alpha\eta))]}{\sqrt{2\alpha \log(K k/(\alpha \eta))}}.
\label{snr20}
\end{eqnarray}
The contribution from this term then goes to zero as $\eta\rightarrow 0$.
Hence, the entire contribution to the limit of ${\cal I}_{j y}(n,K,k,\eta)$ comes from the lower limit of integration:
\begin{equation}
\lim_{\eta\rightarrow 0} {\cal I}_{j y}(n,K,k,\eta)=\frac{ \left(\frac{K}{k}\right)^{n+1/2}}{ \left(k^2-K^2\right)}.
\label{snr21}
\end{equation}

The third integral we consider is 
\begin{equation}
{\cal I}_{yy}(n,K,k,\eta) = \int_0^\infty x^2 \exp(-\eta x^2) y_{n}(K x) y_{n}(k x) d x.
\label{snr22}
\end{equation}
Using equation (\ref{snr14}), this gives
\begin{equation}
{\cal I}_{yy}(n,K,k,\eta) = \int_0^\infty x^2 \exp(-\eta x^2) j_{-n}(K x) j_{-n}(k x) d x=\frac{\pi}{2 \sqrt{K k} }\frac{ \exp[-(K^2+k^2)/4 \eta]}
{2 \eta} I_{-n-1/2}\left(\frac{K k}{2 \eta}\right).
\label{snr23}
\end{equation}
Note however that the result (\ref{snr23}) as it stands is only valid for $n=0$, as the integrand in (\ref{snr22}) diverges at the  lower limit
in non-integrable fashion for $n\ge 1$.

The equation (\ref{snr16}) also holds if $b$ is replaced by $-b$ (the leading term in the equations  (\ref{nr17}-\ref{nr19}) being unaffected by this change). Hence, for $n=0$,
\begin{equation}
\lim_{\eta\rightarrow 0}  {\cal I}_{y y}(0,K,k,\eta)=\frac{\pi}{2 K k} \delta (K-k).
\label{snr24}
\end{equation}
For $n\neq 0$, we have to deal with the lower limit of the integral appropriately (see below).

Other integral evaluations also follow from equations (\ref{snr13},\ref{snr14}). The first of these comes from $ {\cal I}_{j j}(n,K,k,\eta)$, and so is symmetric under interchange of $K$ and $k$: with
\begin{equation}
{\cal I}_{jym}(n,K,k,\eta) = \int_0^\infty x^2 \exp(-\eta x^2) j_{n}(K x) \sqrt{\frac{\pi}{2 x}} Y_{-n-1/2}(k x) d x,
\label{snr25}
\end{equation}
then
\begin{equation}
{\cal I}_{jym}(n,K,k,\eta) ={\cal I}_{jym}(n,k,K,\eta) =(-1)^n {\cal I}_{j j}(n,K,k,\eta).
\label{snr26}
\end{equation}
This integral converges for all $n>0$.
A similar identity exists for the integral ${\cal I}_{jym}(n,K,k,\eta)$, but is not of interest since the integral diverges at its lower limit, even for $n=0$. Other identities start from  $ {\cal I}_{j y}(n,K,k,\eta)$, and so  will not be symmetric under interchange of $K$ and $k$. These
identify ${\cal I}_{jym}(n,K,k,\eta)$ with   $(-1)^n {\cal I}_{ym y}(n,K,k,\eta)$ and $(-1)^{n+1} {\cal I}_{j jm}(n,K,k,\eta)$.
\subsection{Numerical examples}
We now give some examples of spherical Bessel product integrals with Gaussian factors, In each case the analytic formulae have been compared with results obtained by numerical integration in Mathematica, with agreement to all digits quoted in the Table \ref{table1}. A strongly-localising Gaussian has been used in the examples given, so as to facilitate comparisons with other techniques if desired. Of course, more stringent tests could be achieved by diminishing the value of $\eta$; this would necessitate integrating to larger values of $x$ to achieve a satisfactory "killing" of the oscillating integrand.

\begin{table}
\begin{tabular}{|c|c|c|c|c|c|}\\ \hline
$n$& $K$& $k$&$\eta$ & ${\cal I}$ &Value\\ \hline
2 & 1.37 & 2.96 &3.58 & ${\cal I}_{jj}$ & 0.000680896 \\
2 & 1.37+0.457 $i$ & 2.96+1.479 $i$  &3.58 & ${\cal I}_{jj}$ & 0.000741033 + 0.00100379  $i$ \\
3 & 1.37 & 2.96 &3.58 & ${\cal I}_{jj}$ & 0.000054813 \\
3 & 1.37+0.457 $i$ & 2.96+1.479 $i$  &3.58 & ${\cal I}_{jj}$ & -0.0000260529 + 0.000120958 $i$ \\ \hline
0 & 1.37 & 2.96 &3.58 & ${\cal I}_{yy}$ & 0.0639986 \\
0 & 1.37+0.457 $i$ & 2.96+1.479 $i$  &3.58 & ${\cal I}_{yy}$ & 0.00806694 - 0.0549797 $i$ \\ \hline
0& 1.37 & 2.96 &3.58 & ${\cal I}_{jy}$ & -0.00941848 \\
0 & 1.37+0.457 $i$ & 2.96+1.479 $i$  &3.58 & ${\cal I}_{jy}$ & -0.00948972 + 0.00346762 $i$ \\
1 & 1.37 & 2.96 &3.58 & ${\cal I}_{jy}$ & -0.00851273 \\
1 & 1.37+0.457 $i$ & 2.96+1.479 $i$  &3.58 & ${\cal I}_{jy}$ & -0.00586463 + 0.00505498 $i$ \\
3 & 1.37 & 2.96 &3.58 & ${\cal I}_{jy}$ & -0.000878441 \\
3 & 1.37+0.457 $i$ & 2.96+1.479 $i$  &3.58 & ${\cal I}_{jy}$ &-0.000336487 + 0.000656101  $i$ \\
\hline
\end{tabular}
\label{table1}
\caption{Numerical examples of spherical Bessel function values, evaluated both by the analytic formulae of this section and by numerical integration}
\end{table}

\begin{figure}[tbh]
\includegraphics[width=6cm]{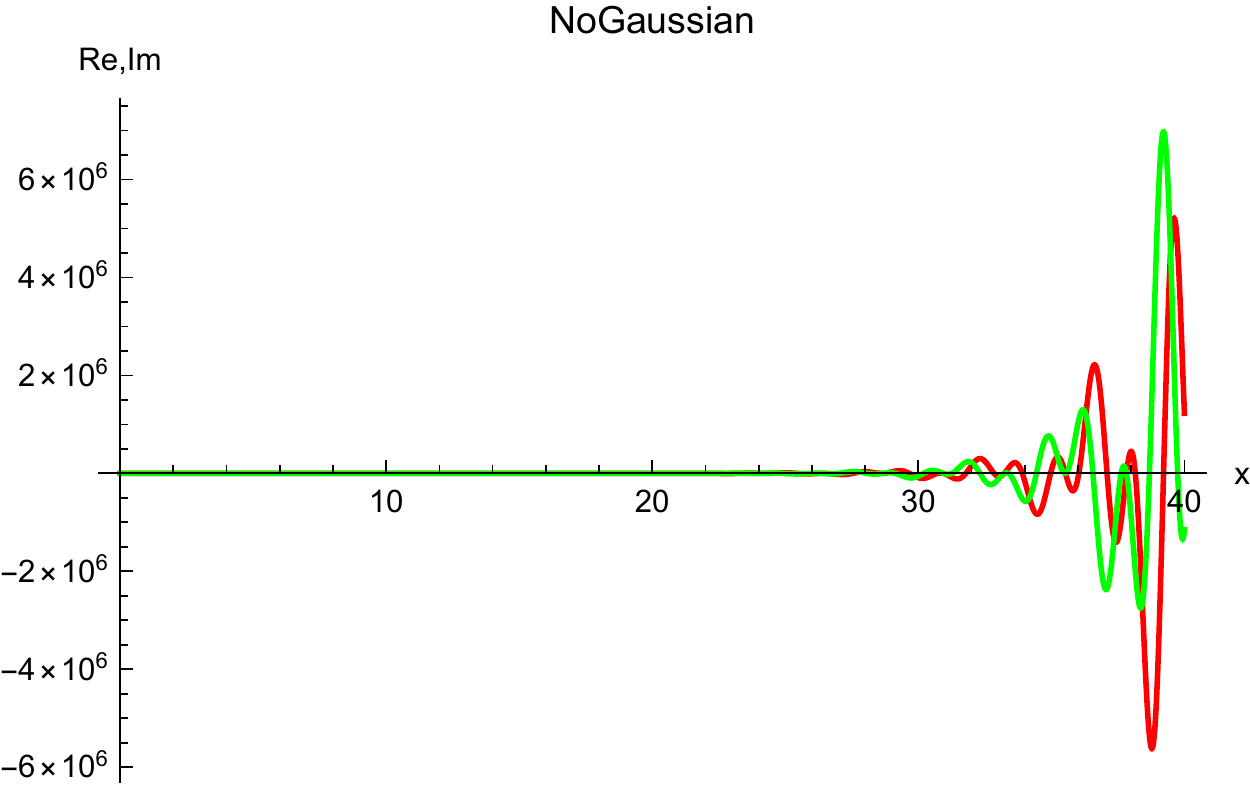}~~\includegraphics[width=6cm]{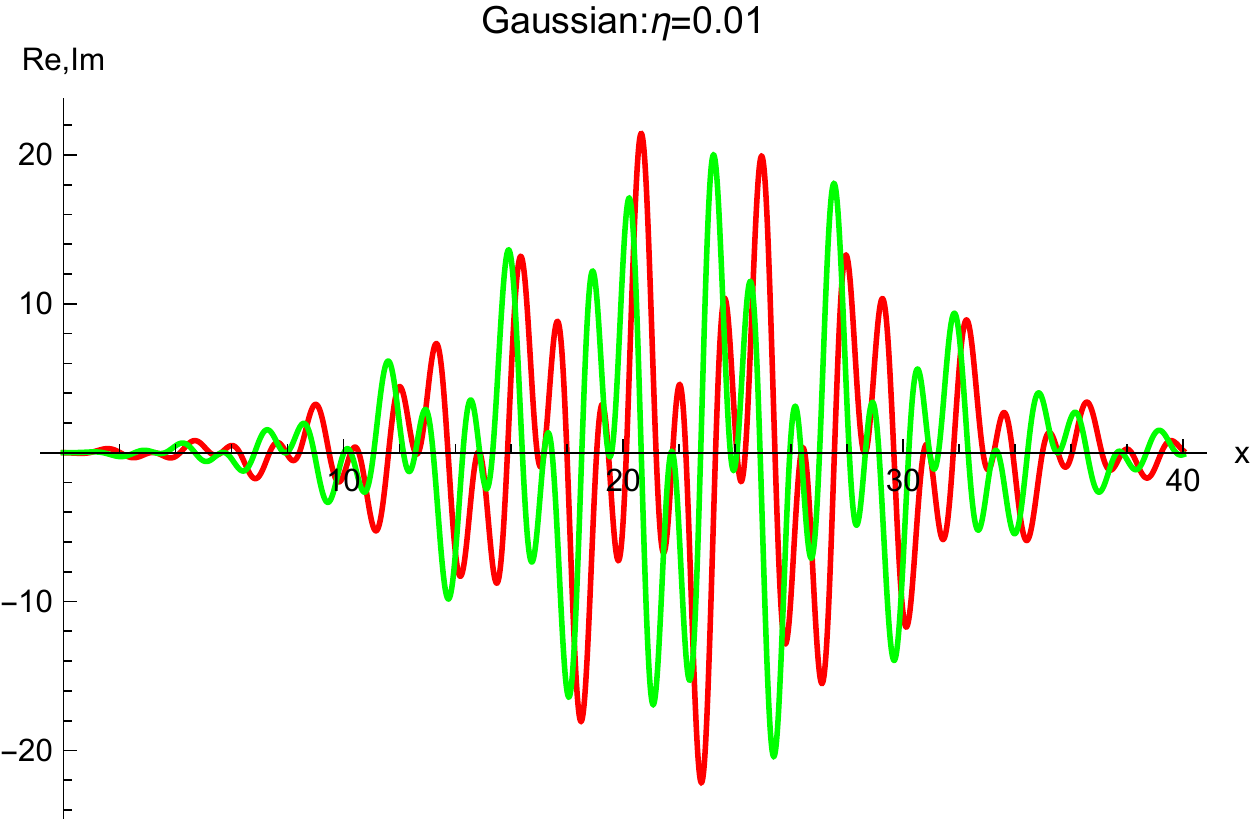}
\caption{The effect of a Gaussian "killing" function on ${\cal I}_{jy}$: $n=1$, $k=1.37$, $K=2.96 + 0.457 i$. At left: without the Gaussian factor, the integrand diverges strongly; at right, with the Gaussian factor with $\eta=0.01$ the  integrand converges to zero for large $x$.}
\label{figkill}
\end{figure}

We now give an example of the effectiveness of the Gaussian "killing" function technique: see Fig. \ref{figkill}. Even with a Gaussian with $\eta$ only equal to 0.01, the divergent integrand is replaced by one which can be integrated accurately.
The numerical integration of the Gaussian form with $\eta =0.01$ gives $0.0164787 - 0.0138487 i$,
for integration with upper limit 80 or beyond. 
The analytic value for the integral (\ref{watsoninteg})  with the upper limit set to infinity  is
$0.0163332 - 0.0135188 i$. 
For $\eta=0.005$, the numerical integral gives $0.0164062 - 0.0136812 i$,  slightly closer to the exact answer, while
for $\eta=0.001$,  the numerical integration in Mathematica fails .

\subsection{Integrals from $R>0$}
In the modes associated with the scattering by  spheres of a given radius, $R$ say, integrals over all space involve integrands  differing  in $r<R$ and 
$r>R$, with the former being in general non-singular at the origin. The integrals over  $r<R$ can be evaluated  using the integral already referred to from 
Watson, p.134:
\begin{equation}
\int^{z} z {\cal C}_\mu (k z)  {\cal D}_\mu (l z) d z= \frac{z\left\{k {\cal C}_{\mu+1} (k z) {\cal D}_\mu (l z) -l {\cal C}_\mu (k z)  
{\cal D}_{\mu+1} (l z)\right\} }{k^2-l^2},
\label{watsoninteg}
\end{equation}
where ${\cal C}$ and ${\cal D}$ are cylinder functions.
The integrals over $r>R$ can be dealt with using the same integral:
\begin{equation}
\int_R^\infty r^2 {\cal C}_n(K r){\cal D}_n(k r) d r=\lim_{\delta \rightarrow 0} [ \int_\delta^\infty r^2 {\cal C}_n(K r){\cal D}_n(k r) d r
- \int_\delta^R r^2 {\cal C}_n(K r){\cal D}_n(k r) d r].
\label{snr27}
\end{equation}
Both the integrals on the right-hand side are well behaved or have the same singularity at $\delta=0$, which can be canceled.
The problem arising at the upper limit of the first integral has been dealt with already using  the limit of Gaussians, so that we know how to  deal with 
both contributions to the integral appropriately.
 
 We thus arrive at the key results of this paper in relation to spherical Bessel integrals:
 \begin{equation}
 \int_{R}^\infty x^2 j_n(K x) j_n (k x)d x=\frac{\pi}{2 K k} \delta (K-k)- \frac{R^2\left\{K {j}_{n+1} (K R) {j}_n (k R) -k {j}_n (K R)  {j}_{n+1} (k R)
 \right\} }{K^2-k^2},
 \label{key1}
 \end{equation}
  \begin{equation}
 \int_{R}^\infty x^2 j_n(K x) y_n (k x)d x=- \frac{R^2\left\{K {j}_{n+1} (K R) {y}_n (k R) -k {j}_n (K R)  {y}_{n+1} (k R)\right\} }{K^2-k^2},
 \label{key2}
 \end{equation}
 and
\begin{equation}
 \int_{R}^\infty x^2 y_n(K x) y_n (k x)d x=\frac{\pi}{2 K k} \delta (K-k)- \frac{R^2\left\{K {y}_{n+1} (K R) {y}_n (k R) -k {y}_n (K R)  {y}_{n+1} (k R)\right\} }{K^2-k^2}.
 \label{key2a}
\end{equation}

\section{Spherical Bessel function product integrals  for resonant state calculations}
\subsection{Bessel function integrals for magnetic source E-fields}
\label{BessInt}

\subsubsection{Regular spherical Bessel function integrals}

The integrals we want to calculate are of the type,
\begin{equation}
\int d^{3}r\boldsymbol{M}_{n,m}\left(  K\boldsymbol{r}\right) \cdot
\boldsymbol{M}_{n,m}\left(  k\boldsymbol{r}\right)  
 =\int x^{2}j_{n}\left(  Kx\right)  j_{n}\left(  kx\right)  dx\;. \label{Mjint}
\end{equation}

One way to write this Bessel function integral with an infinite upper limit is,
\begin{equation}
\int_{R}^{\infty}x^{2}j_{n}\left(  Kx\right)  j_{n}\left(  kx\right)
dx=\frac{\pi}{2Kk}\delta\left(  K-k\right)  + R^{2}\frac{ Kj_{n}
\left(  kx\right)  j_{n+1}\left(  Kx\right)  -kj_{n}\left(  Kx\right)  j_{n+1}
\left(  kx\right) }{K^{2}-k^{2}}\;, \label{Ross1}
\end{equation}
Alternative analytic expressions for eq.(\ref{Mjint}) entirely in terms of Bessel functions of the same order
$n$ as the integrand are be obtained by invoking Bessel function derivatives. For finite intervals these expressions are,
\begin{subequations}
\begin{align}
\int_{0}^{R}x^{2}j_{n}\left(  Kx\right)  j_{n}\left(  kx\right)  dx
&=\frac{k \psi_{n}^{\prime}\left(  kR\right) \psi_{n}\left(  KR\right) 
-K \psi_{n}^{\prime}\left(  KR\right) \psi_{n}\left(  kR\right)  }{Kk\left( K^{2}-k^{2} \right) }
\label{0toR} \\
\begin{split}
\int_{R}^{L}x^{2}j_{n}\left(  Kx\right)  j_{n}\left(  kx\right)  dx & =  
\frac{k \psi_{n}^{\prime}\left(  kL\right) \psi_{n}\left(  KL\right)  
-K \psi_{n}^{\prime}\left(  KL\right)\psi_{n}\left(  kL\right)  }{Kk\left( K^{2}-k^{2} \right) }\\
& \qquad -\frac{k \psi_{n}^{\prime}\left(  kR\right)\psi_{n}\left(  KR\right) 
  -K \psi_{n}^{\prime}\left(  KR\right)\psi_{n}\left(  kR\right)  }{Kk\left( K^{2}-k^{2} \right) }
\;,\label{RtoL}
\end{split}
\end{align}
\end{subequations}
while for the infinite interval one obtains,
\begin{equation}
\int_{R}^{\infty}x^{2}j_{n}\left(  Kx\right)  j_{n}\left(  kx\right)
dx=\frac{\pi}{2Kk}\delta\left(  K-k\right) -\frac{k \psi_{n}^{\prime}\left(  kR\right)\psi_{n}\left(  KR\right) 
  -K \psi_{n}^{\prime}\left(  KR\right)\psi_{n}\left(  kR\right)  }{Kk\left( K^{2}-k^{2} \right) }\;. \label{Br1}
\end{equation}\label{Brform}
Bessel function recurrence relations readily show that eqs.(\ref{Ross1}) and
(\ref{Br1}) are just alternate expressions of the same quantity.

Special attention needs to be paid to the case $K=k$,  where eqs.(\ref{Ross1}) and (\ref{Brform}) encounter problems.
The following alternative expression may be established:
\begin{align}
\begin{split}
&  \underset{K\rightarrow k}{\lim} \frac{k \psi_{n}^{\prime}\left(  kR\right) \psi_{n}\left(  KR\right) 
-K \psi_{n}^{\prime}\left(  KR\right) \psi_{n}\left(  kR\right)  }{Kk\left( K^{2}-k^{2} \right)}\\
&  =\frac{R}{2}\frac{\left[  \psi_{n}^{\prime}\left(  kR\right)  \right]^{2}
+\psi_{n}^{2}\left(  kR\right)  -n\left(  n+1\right)  j_{n}^{2} \left(
kR\right)  -j_{n}\left(  kR\right)  \psi_{n}^{\prime}\left(  kR\right)  }
{k^{2}}\;,\label{Ktoklim}
\end{split}
\end{align}
where $\psi_{n}\left(  x\right)  \equiv xj_{n}\left(  x\right)  $ are the
Ricatti Bessel functions. This result gives the following definite integrals,
\begin{subequations}
\begin{align}
\int_{0}^{R}x^{2}j_{n}^{2}\left(  kx\right)  dx  &  =\frac{R}{2}\frac{\left[
\psi_{n}^{\prime}\left(  kR\right)  \right]  ^{2}+\psi_{n}^{2}\left(
kR\right)  -n\left(  n+1\right)  j_{n}^{2}\left(  kR\right)  -j_{n} \left(
kR\right)  \psi_{n}^{\prime}\left(  kR\right)  }{k^{2}}\label{Min}\\
\begin{split}
\int_{R}^{L}x^{2}j_{n}^{2}\left(  kx\right)  dx  &  =\left\{  \frac{L}{2}
\frac{\left[  \psi_{n}^{\prime}\left(  kL\right)  \right]  ^{2} +\psi_{n}^{2}
\left(  kL\right)  -n\left(  n+1\right)  j_{n}^{2}\left(  kL\right)
-j_{n}\left(  kL\right)  \psi_{n}^{\prime}\left(  kL\right)  }{k^{2}}\right.
\\
&  \left. \qquad -\frac{R}{2}\frac{\left[  \psi_{n}^{\prime}\left(  kR\right)
\right]  ^{2}+\psi_{n}^{2}\left(  kR\right)  -n\left(  n+1\right)  j_{n}^{2}
\left(  kR\right)  -j_{n}\left(  kR\right)  \psi_{n}^{\prime}
\left(kR\right)  } {k^{2}}\right\}  \;,\label{Mext}
\end{split}
\end{align}
\end{subequations}

A comparison of eqs.(\ref{Br1}), (\ref{Ktoklim}) and (\ref{Mext}) above allows
us to conclude that,
\begin{align}
\frac{\pi}{2Kk}\delta\left(  K-k\right)  = \underset{L\rightarrow\infty}{\lim}
\frac{k \psi_{n}^{\prime}\left(  kL\right) \psi_{n}\left(  KL\right) 
-K \psi_{n}^{\prime}\left(  KL\right) \psi_{n}\left(  kL\right)  }{Kk\left( K^{2}-k^{2} \right)} \;. \label{delta_rep_J}
\end{align}
The veracity of eq.(\ref{delta_rep_J}) may be verified   by the  method of Section 3.
An important special case of the above integrals is the orthogonality of the 
free-space wave functions,
\begin{equation}
\int_{0}^{\infty}dxx^{2}j_{n}\left(  Kx\right)  j_{n}\left(  kx\right)
=\frac{\pi}{2Kk}\delta\left(  K-k\right)  \;.
\end{equation}

\subsubsection{Spherical Neumann integrals}

The spherical Neumann functions are ill-defined when their argument goes to
zero, but one can still evaluate integrals not involving the origin. Notably, one
has,
\begin{align}
\int_{R}^{\infty}x^{2}y_{n}\left(  Kx\right)  y_{n}\left(  kx\right)
dx=\frac{\pi}{2Kk}\delta\left(  K-k\right)  +R^{2}\left[\frac{
Ky_{n+1}\left(  Kx\right)  y_{n}\left(  kx\right)  -ky_{n}\left(  Kx\right)
y_{n+1}\left(  kx\right) }{K^{2}-k^{2}}\right]\;, \label{Rossyint}
\end{align}
while the expression for the definite integral using Neumann function derivatives is,
\begin{align}
\begin{split}
\int_{R}^{L}x^{2}y_{n}\left(  Kx\right)  y_{n}\left(  kx\right)  dx  &
=\frac{k\chi_{n}^{\prime}\left(  kL\right) \chi_{n}\left(  KL\right)  
-K \chi_{n}^{\prime}\left(  KL\right) \chi_{n}\left(  kL\right)}{K k \left( K^{2}-k^{2} \right)}\\
&  \qquad-R^{2}\frac{k \chi_{n}^{\prime}\left(  kR\right) \chi_{n}\left(  KR\right)  
-K \chi_{n}^{\prime}\left(  KR\right) \chi_{n}\left(  kR\right)  }{K k \left( K^{2}-k^{2} \right)}\;.
\end{split}\label{Bryfinint}
\end{align}
Arguing in analogy with eq.(\ref{Ktoklim}) that,
\begin{align}
\frac{\pi}{2Kk}\delta\left(  K-k\right)  =\underset{L\rightarrow\infty}{\lim}
\frac{k \chi_{n}^{\prime}\left(  kL\right) \chi_{n}\left(  KL\right)  
-K \chi_{n}^{\prime}\left(  KL\right) \chi_{n}\left(  kL\right)    }{K k \left( K^{2}-k^{2} \right)}\;,
\end{align}
the $L\rightarrow\infty$ limit of  eq.(\ref{Bryfinint}) then reads,
\begin{align}
\int_{R}^{\infty}x^{2}\chi_{n}\left(  Kx\right) \chi_{n}\left(  kx\right)
dx=\frac{\pi}{2Kk}\delta\left(  K-k\right)  - \frac{k \chi_{n}
\left(KR\right) \chi_{n}^{\prime}\left(  kR\right)  -K \chi_{n}\left(  kR\right)
\chi_{n}^{\prime}\left(  KR\right)}{K k \left( K^{2}-k^{2} \right)}\;,
\end{align}
which one can verify is just another way of writing eq.(\ref{Rossyint}).

One can again evaluate the $K\rightarrow k$ of the second term on the right hand
side of eq.(\ref{Bryfinint}) with the expression,
\begin{align}
\begin{split}
&  \underset{K\rightarrow k}{\lim}\frac{  k\chi_{n}\left(  KR\right)
\chi_{n}^{\prime}\left(  kR\right)  -K \chi_{n}\left(  kR\right)  \chi_{n}^{\prime}
\left(  KR\right)   }{K k \left( K^{2}-k^{2} \right)}\\
&  \qquad=\frac{R}{2}\frac{\left[  \chi_{n}^{\prime}\left(  kR\right)
\right]^{2} +\chi_{n}^{2}\left(  kR\right)  -n\left(  n+1\right)  
y_{n}^{2}\left(  kR\right)  -y_{n}\left(  kR\right)  \chi_{n}^{\prime}
\left(kR\right) }{k^{2}}\;,
\end{split}
\end{align}
where $\chi_{n}(z)=xy_{n}(z)$ are Ricatti Neumann functions.

Finite integrals of the Neumann functions squared are then,
\begin{align}\begin{split}
\int_{R}^{L}x^{2}y_{n}^{2}\left(  kx\right)  dx  &  =\left\{  \frac{L}{2}
\frac{\left[  \chi_{n}^{\prime}\left(  kL\right)\right]^{2}+\chi_{n}^{2}
\left(  kL\right)  -n\left(  n+1\right)  y_{n}^{2}\left(  kL\right)
-y_{n}\left(  kL\right)  \chi_{n}^{\prime}\left(  kL\right)  }{k^{2}}\right.
\\
& \qquad\qquad \left.  -\frac{R}{2}\frac{\left[  \chi_{n}^{\prime}\left(  kR\right)\right]^{2}
+\chi_{n}^{2}\left(  kR\right)  -n\left(  n+1\right)  
y_{n}^{2}\left(  kR\right)  -y_{n}\left(  kR\right)  \psi_{n}^{\prime} 
\left(kR\right)  }{k^{2}}\right\}  \;.
\end{split}\end{align}
Another remark is that all the finite domain integrals hold even if $K$ and
$k$ are complex valued. 

\subsubsection{Mixed Bessel-Neumann integrals}

For the mixed Bessel-Neumann integrals one can extend the lower bound to zero and we have,
\begin{align}\begin{split}
\int_{0}^{L}x^{2}j_{n}\left(  Kx\right)  y_{n}\left(  kx\right)  dx &
=\frac{k \psi_{n}\left(  KL\right)  \chi_{n}^{\prime}\left(  kL\right)
-K\psi_{n}^{\prime}\left(  KL\right) \chi_{n}\left(  kL\right)}{K k \left( K^{2}-k^{2} \right)} 
-\frac{K^{n}}{k^{n+1}\left(  K^{2}-k^{2}\right)  } \;,
\end{split}\end{align}
while the indefinite integrals are,
\begin{align}\begin{split}
\int_{R}^{\infty} x^{2} j_{n}\left(  Kx\right)  y_{n}\left(  kx\right)
dx & =-\frac{k\psi_{n}\left(  KR\right)  \chi_{n}^{\prime}\left(  kR\right)-K\psi_{n}^{\prime}
\left(  KR\right) \chi_{n}\left(  kR\right)}{K k \left( K^{2}-k^{2} \right)} \;, \\
\int_{0}^{\infty} j_{n}\left(  Kx\right)  y_{n}\left(  kx\right)
dx  &= -\frac{K^{n}}{k^{n+1}\left(  K^{2}-k^{2}\right)  } 
\end{split}\end{align}
where the lack of a delta-function contribution is important.

We can also obtain the results for limits of $K\rightarrow k$,
\begin{align}\begin{split}
&\int_{0}^{L}x^{2}j_{n}\left(  kx\right)  y_{n}\left(  kx\right)  dx= \\
& \frac{\psi_{n}^{\prime}\left(  kL\right)\chi_{n}^{\prime}\left(  kL\right) 
+\psi_{n}\left(  kL\right)\chi_{n}\left(  kL\right)  -n\left(  n+1\right) 
y_{n}\left(  kL\right)j_{n}\left(  kL\right)
- \psi_{n}^{\prime}\left(  kL\right) y_{n}\left(  kL\right)  }{2k^{2}/L} - \frac{n+1}{2k^{3}} \;.
\end{split}\end{align}
with a particularly simple result for the indefinite integral,
\begin{align}\begin{split}
&\int_{0}^{\infty}x^{2}j_{n}\left(  kx\right)  y_{n}\left(  kx\right)  dx=  - \frac{n+1}{2k^{3}} \;.
\end{split}\end{align}

\subsubsection{Spherical Hankel function integrals}

Recalling that the spherical Hankel functions are defined,
\begin{equation}
h_{n}\left(  x\right)  \equiv j_{n}\left(  x\right)  +iy_{n}\left(  x\right) \;,
\end{equation}
this means,
\begin{align}
h_{n}\left(  Kx\right)  h_{n}\left(  kx\right)  =j_{n}\left(  Kx\right)j_{n}\left(  kx\right)  
-y_{n}\left(  Kx\right) y_{n}\left(  kx\right) + i \left\{  j_{n}\left(  Kx\right)y_{n}\left(  kx\right) 
+ y_{n}\left(  Kx\right)j_{n}\left(  kx\right) \right\} \;.
\end{align}
Consequently, regardless of our interpretation of the delta functions,
$\delta\left(  K-k\right)$, with complex values of $K$ and $k$,
they are going to cancel for the Hankel function integrals to leave us with,

\begin{subequations}
\begin{align}
\int_{R}^{\infty}x^{2}h_{n}\left(  Kx\right)  h_{n}\left(  kx\right)  dx  &
=-\frac{k\xi_{n}\left(  KR\right)  \xi_{n}^{\prime}\left(  kR\right)
-K\xi_{n}\left(  kR\right) \xi_{n}^{\prime}\left(  KR\right)  }{K k \left( K^{2}-k^{2} \right)}\;\\
\int_{R}^{\infty}x^{2}h_{n}^{2}\left(  kx\right)  dx  & =-\frac{R}{2}
\frac{\left[  \xi_{n}^{\prime}\left(  kR\right)  \right]  ^{2}
+\xi_{n}^{2}\left(  kR\right)  -n\left(  n+1\right)  h_{n}^{2}\left(  kR\right)
-h_{n}\left(  kR\right)  \xi_{n}^{\prime}\left(  kR\right)  }{k^{2}} \;,
\end{align}
\end{subequations}
which are precisely the results needed for normalization and
orthogonalization.

\subsection{Bessel function integrals for electric type fields}

\subsubsection{Bessel and Neumann product integrals}

For electric type fields, the integrals that one needs to evaluate are (for regular fields),
\begin{align}\begin{split}
& \int d^{3}r \boldsymbol{N}_{n,m}\left(  K\boldsymbol{r}\right) \cdot
\boldsymbol{N}_{n,m}\left(  k \boldsymbol{r}\right)  \\
& \qquad =\int \frac{  n\left(  n+1\right)  j_{n}\left(  Kx\right)
j_{n}\left(  kx\right)  +\psi_{n}^{\prime}\left(  Kx\right)  \psi_{n}^{\prime}
\left(  kx\right) }{Kk}  dx\;,
\end{split}\end{align}
but also carry out the analogous integrals for outgoing partial waves with
the Bessel $j_n$ functions replaced by outgoing Hankel functions, $h_n$, and the 
$\psi_{n}(z)=zj_n(z)$  replaced by $\xi_{n}(z)=zh_n(z)$.

For finite integrals  with $K\neq k$ one finds,
\begin{align}\begin{split}
& \int_{R}^{L}\frac{ n\left(  n+1\right)  j_{n}
\left(Kx\right)  j_{n}\left(  kx\right)  +\psi_{n}^{\prime}\left(  Kx\right)
\psi_{n}^{\prime}\left(  kx\right)}{Kk}  dx\\
& \qquad \qquad =\frac{K\psi_{n}\left(  KL\right)  \psi_{n}^{\prime}\left(
kL\right)  -k\psi_{n}\left(  kL\right)  \psi_{n}^{\prime}\left(  KL\right)}
{Kk \left( K^{2}-k^{2} \right)}\\
& \qquad \qquad\qquad  -\frac{K \psi_{n}\left(  KR\right)  \psi_{n}^{\prime}
\left(kR\right)  -k \psi_{n}\left(  kR\right)  \psi_{n}^{\prime}\left(  KR\right)}
{ Kk\left( K^{2}-k^{2} \right) } \;, \label{Njfin}
\end{split}\end{align}
\begin{align}\begin{split}
& \int_{0}^{R} \frac{  n\left(  n+1\right)  j_{n}\left(
Kx\right)  j_{n}\left(  kx\right)  +\psi_{n}^{\prime}\left(  Kx\right)
\psi_{n}^{\prime}\left(  kx\right) }{Kk} dx \\
& \qquad \qquad =\frac{K \psi_{n}\left(  KR\right)  \psi_{n}^{\prime}
\left(kR\right)  -k \psi_{n}\left(  kR\right)  \psi_{n}^{\prime}\left(  KR\right)}
{K k \left( K^{2}-k^{2}\right) }\;.
\end{split}\end{align}
A finite integral for $K=k$ is,
\begin{align}\begin{split}
& \int_{R}^{L} \frac{  n\left(  n+1\right)  j_{n}^{2}\left(  kx\right) 
+\left[\psi_{n}^{\prime}\left(  kx\right)\right]^{2} }{k^{2}}  dx\\
& \qquad \qquad=\frac{L}{2}\frac{\left[  \psi_{n}^{\prime}\left(  kL\right)\right]^{2}
+\psi_{n}^{2}\left(  kL\right)  -n\left(  n+1\right)  j_{n}^{2}\left(kL\right)  
+j_{n}\left(  kL\right)  \psi_{n}^{\prime}\left(  kL\right)}{k^{2}}\\
&\qquad \qquad  \qquad-\frac{R}{2}\frac{\left[  \psi_{n}^{\prime}\left(  kR\right) 
 \right]^{2}+\psi_{n}^{2}\left(  kR\right)  -n\left(  n+1\right)  j_{n}^{2}
\left(kR\right)  +j_{n}\left(  kR\right)  \psi_{n}^{\prime}\left(  kR\right)}{k^{2}} \;.
\end{split}\end{align}

In the $L\rightarrow\infty$ limit, the first term on the right hand side of the last line
vanishes, while the second term yields a delta function once again through,
\begin{align}
\frac{\pi}{2Kk}\delta\left(  K-k\right)  = \underset{L\rightarrow\infty}{\lim}
\frac{K\psi_{n}\left(  KL\right)  \psi_{n}^{\prime}\left(
kL\right)  -k\psi_{n}\left(  kL\right)  \psi_{n}^{\prime}\left(  KL\right)}
{Kk \left( K^{2}-k^{2} \right)} \;.
\end{align}
and we arrive at the required indefninte integral,
\begin{align}\begin{split}
& \int_{R}^{\infty} \frac{  n\left(  n+1\right)  j_{n}\left(
Kx\right)  j_{n}\left(  kx\right)  +\psi_{n}^{\prime}\left(  Kx\right)
\psi_{n}^{\prime}\left(  kx\right)}{Kk}  dx \\
&  =\frac{\pi}{2Kk}\delta\left(  K-k\right)  -\frac{K \psi_{n}\left(  KR\right)  
\psi_{n}^{\prime}\left(  kR\right)  -k \psi_{n}
\left(kR\right)  \psi_{n}^{\prime}\left(  KR\right)}{Kk\left( K^{2}-k^{2} \right) }\;.
\end{split}\end{align}

The same procedure for $y_{n}$ product integrals yields,
\begin{align}\begin{split}
& \int_{R}^{\infty} \frac{ n\left(  n+1\right)  y_{n}\left(
Kx\right)  y_{n}\left(  kx\right)  +\chi_{n}^{\prime}\left(  Kx\right)
\chi_{n}^{\prime}\left(  kx\right) }{Kk}  dx\\
&  =\frac{\pi}{2Kk}\delta\left(  K-k\right)  -\frac{K \chi_{n}\left(  KR\right)  
\chi_{n}^{\prime}\left(  kR\right)  -k \chi_{n}
\left(kR\right)  \chi_{n}^{\prime}\left(  KR\right)}{Kk\left( K^{2}-k^{2}\right) }\;.
\end{split}\end{align}

\subsection{Hankel product integrals}

The Hankel function integrals that we need for field normalization are,
\begin{subequations}
\begin{align}
\begin{split}
& \int_{R}^{\infty}\frac{ n\left(  n+1\right)  h_{n}
\left(Kx\right)  h_{n}\left(  kx\right)  + \xi_{n}^{\prime}\left(  Kx\right)
\xi_{n}^{\prime}\left(  kx\right)}{Kk}  dx \\
& \qquad \qquad =-\frac{K \xi_{n}\left(  KR\right)  
\xi_{n}^{\prime}\left(  kR\right)-k \xi_{n}\left(  kR\right)  
\xi_{n}^{\prime}\left(  KR\right)}{Kk\left( K^{2}-k^{2} \right)}\; ,
\end{split}\\
 {\rm and} &  \;,\nonumber \\
\begin{split}
 &\int_{R}^{\infty}\frac{  n\left( n+1\right)   \xi_{n}^{2}\left( kx\right)  
+  \left[\xi_{n}^{\prime}\left(  kx\right)\right]^{2} }{k^{2}}  dx  \\ 
&  \qquad \qquad  =-\frac{R}{2}
\frac{\left[  \xi_{n}^{\prime}\left(  kR\right)\right]^{2}
+\xi_{n}^{2}\left(  kR\right)  -n\left(  n+1\right)  \xi_{n}^{2}\left(  kR\right)
+ h_{n}\left(  kR\right)  \xi_{n}^{\prime}\left(kR\right)}{k^{2}} \;.
\end{split}\end{align}\label{hankHinf}
\end{subequations}

\section{Conclusions}
The analytic expressions we have given involving products of Bessel functions combined with a Gaussian term are easily verified numerically, and this has been done for the results given here. The results obtained analytically for integrals with the Gaussian having tended to a constant may also be tested numerically, as was exemplified in Section 5.1. Another test, which will be reported on in a future publication, is provided by residue calculus of field scattering amplitudes, which provides a second route to mode normalisation factors. In fact, these two analytic approaches agree completely.
\section{Acknowedgements}
Research conducted within the context of the International Associated Laboratory for Photonics
between France and Australia. This work has been carried out thanks to the support of the A*MIDEX
project (no. ANR-11-IDEX-0001-02) funded by the Investissements d'Avenir French Government program, managed by the French National Research Agency (ANR). The authors would like to thank Remi Colom, Nicolas Bonod and Thomas Durt for helpful discussions.
\bibliographystyle{unsrt}

\begin{thebibliography}{10}

\bibitem{Mie1908}
Gustav Mie.
\newblock Beiträge zur optik trüber medien, speziell kolloidaler
  metallösungen.
\newblock {\em Annalen der Physik}, 330(3):377--445, 1908.

\bibitem{CalderonPeierls1976}
Gaston García-Calderón and Rudolf Peierls.
\newblock Resonant states and their uses.
\newblock {\em Nuclear Physics A}, 265(3):443 -- 460, 1976.

\bibitem{zeld61}
Ya.~B. Zel'dovich.
\newblock On the theory of unstable states.
\newblock {\em Sov. Phys. JETP}, 12:542--545, 1961.

\bibitem{perezeld}
A.M. Perelomov and Y.B. Zel'dovich.
\newblock {\em Quantum Mechanics: Selected Topics}.
\newblock World Scientific, Singapore, 1998.

\bibitem{Calderon2010}
Gastón García-Calderón.
\newblock Chapter 7 - theory of resonant states: An exact analytical approach
  for open quantum systems.
\newblock In Cleanthes~A. Nicolaides and Erkki Brändas, editors, {\em Unstable
  States in the Continuous Spectra, Part I: Analysis, Concepts, Methods, and
  Results}, volume~60 of {\em Advances in Quantum Chemistry}, pages 407 -- 455.
  Academic Press, 2010.

\bibitem{Dubovik1990}
V.M. Dubovik and V.V. Tugushev.
\newblock Toroid moments in electrodynamics and solid-state physics.
\newblock {\em Physics Reports}, 187(4):145 -- 202, 1990.

\bibitem{Mulj16}
E.~A. Muljarov and W.~Langbein.
\newblock Exact mode volume and Purcell factor of open optical systems.
\newblock {\em Phys. Rev. B}, 94:235438, Dec 2016.

\bibitem{MDS92}
R.~C. McPhedran, D.~H. Dawes, and T.~C. Scott.
\newblock On a Bessel function integral.
\newblock {\em Applicable Algebra in Engineering, Communication and Computing},
  2(3):207--216, Sept 1992.

\bibitem{GR}
I.~S. Gradshteyn and I.~M. Ryzhik.
\newblock {\em Table of Integrals, Series and Products, Sixth Edition}.
\newblock Alan Jeffrey and Dan Zwillinger, Academic Press, 2000.

\bibitem{Peter86}
P.~A. Robinson.
\newblock Relativistic plasma dispersion functions.
\newblock {\em Journal of Mathematical Physics}, 27(5):1206--1214, 1986.

\bibitem{Luke62}
Y.~L. Luke.
\newblock {\em Integrals of Bessel functions}.
\newblock McGraw-Hill, New York, 1962.

\bibitem{NIST}
F.W. Olver, editor.
\newblock {\em NIST Handbook of Mathematical Functions}.
\newblock Cambridge University Press, Cambridge, 2010.

\bibitem{Lekner18}
John Lekner.
\newblock {\em Theory of Electromagnetic Pulses}.
\newblock 2053-2571. Morgan \& Claypool Publishers, 2018.

\bibitem{Watson80}
G.N. Watson.
\newblock {\em A Treatise on the Theory of Bessel Functions}.
\newblock Cambridge University Press, Cambridge, 1980.

\end{thebibliography}

\end{document}